 \documentclass[a4paper,12pt]{amsart}
 \usepackage{amsmath,amsthm,amssymb}
 \usepackage{graphicx}
 \usepackage[all]{xy}
 \usepackage{txfonts}

\theoremstyle{definition}
\newtheorem{theorem}{Theorem}[section]

\newtheorem{definition}[theorem]{\rm Definition}

\newtheorem{example}[theorem]{\rm Example}
\numberwithin{equation}{section}

\begin{document}

\markboth{G. Matsuda, S. Kaji, and H. Ochiai}
{Dual Complex Numbers}

\title{Anti-commutative Dual Complex Numbers \\ and 2D Rigid Transformation}
\date{}

\author{Genki Matsuda}
\address{Kyushu University / JST CREST}
\email{ma212041@math.kyushu-u.ac.jp}
\author{Shizuo Kaji${}^\dag$}
\thanks{\dag Corresponding author}
\address{Yamaguchi University / JST CREST}
\email[Corresponding author]{skaji@yamaguchi-u.ac.jp}
\author{Hiroyuki Ochiai}
\address{Kyushu University / JST CREST}
\email{ochiai@imi.kyushu-u.ac.jp}


\newcommand{\R}{\mathbb{R}}
\newcommand{\C}{\mathbb{C}}
\renewcommand{\H}{\mathbb{H}}
\newcommand{\GL}{\mathrm{GL}^+}
\newcommand{\Aff}{\mathrm{Aff}^+}
\newcommand{\SO}{\mathrm{SO}}
\newcommand{\E}{\mathrm{E}}
\newcommand{\SE}{\mathrm{SE}}
\newcommand{\se}{\mathfrak{se}}
\newcommand{\Sym}{\mathrm{Sym}^+}
\newcommand{\so}{\mathfrak{so}}
\newcommand{\sym}{\mathfrak{sym}}
\newcommand{\dcn}{\mathfrak{dcn}}
\renewcommand{\Re}{\mathrm{Re}}
\renewcommand{\Im}{\mathrm{Im}}
\newcommand{\T}{{}^t\!}

\maketitle

\begin{abstract}We introduce a new presentation of the two dimensional rigid transformation
which is more concise and efficient than the standard matrix presentation.
By modifying the ordinary dual number construction for the complex numbers,
we define the ring of the \textit{anti-commutative dual complex numbers},
which parametrizes two dimensional rotation and translation all together.
With this presentation, one can easily interpolate or blend 
two or more rigid transformations at a low computational cost.
We developed a library for C++ with the MIT-licensed source code (\cite{source})
and demonstrate its facility by an interactive deformation tool developed for iPad.
\end{abstract}

\section{Rigid Transformation}
\label{sec:1}
The $n$-dimensional {\em rigid transformation (or Euclidean)} group $\E(n)$ consists of
transformations of $\R^n$ which preserves the standard metric.
This group serves as an essential mathematical backend for many applications (see \cite{AO1,AO2}).
It is well-known (see \cite{rigid}, for example) that any element of $\E(n)$
can be written as a composition of a rotation, a reflection, and a translation,
and hence, it is represented by $(n+1)\times (n+1)$-homogeneous matrix;
\[
 \E(n)= \left\{ A=
 \begin{pmatrix} \hat{A} & d_A\\
  0 & 1 
 \end{pmatrix} \mid \hat{A}\T\hat{A}=I_n, d_A\in \R^n
 \right\}.
\]
Here, we adopt the convention that a matrix acts on a (column) vector by the multiplication from the left.
$\E(n)$ has two connected components. The identity component $\SE(n)$
consists of those without reflection. More precisely,
\[
 \SE(n)= \left\{ A=
 \begin{pmatrix} \hat{A} & d_A\\
  0 & 1 
 \end{pmatrix} \mid \hat{A}\in \SO(n), d_A\in \R^n
 \right\},
\]
where $\SO(n)=\{ R \mid R \T R=I_n, \det(R)=1 \}$ is the {\em special orthogonal group}
composed of $n$-dimensional rotations.

The group $\SE(n)$ is widely used in computer graphics such as for
expressing motion and attitude, displacement (\cite{arcball}),
deformation (\cite{arap,igarashi,MLS}), skinning (\cite{DLB}), and camera control (\cite{camera}).
In some cases, the matrix representation of the group $\SE(n)$ is not convenient.
In particular, a linear combination of two matrices in $\SE(n)$ does not necessarily 
belong to $\SE(n)$ and it causes the notorious {\em candy-wrapper} defect in skinning.
When $n=3$, another representation of $\SE(3)$ using the {\em dual quaternion numbers} (DQN, for short) is considered
in \cite{DLB} to solve the candy-wrapper defect.
In this paper, we consider the $2$-dimensional case. 
Of course, 2D case can be handled by regarding
 the plane embedded in $\R^3$ and using DQN,
 but instead, we introduce the {\em anti-commutative dual complex numbers} (DCN, for short),
 which is specific to 2D with much more concise and faster implementation (\cite{source}).
To summarise, our DCN has the following advantages:
\begin{itemize}
\item any number of rigid transformations can be blended/interpolated easily using its algebraic structure
with no degeneration defects such as the candy-wrapper defect (see \S \ref{blend})
\item it is efficient in terms of both memory and CPU usage (see \S \ref{implementation}).
\end{itemize}
We believe that our DCN offers a choice for representing the 2D rigid transformation in certain applications
which requires the above properties.

\section{Anti-commutative dual complex numbers}
Let $\mathbb{K}$ denote one of the fields $\R, \C$, or $\H$, where $\H$ is the quaternion numbers.
First, we recall the standard construction of the {\em dual numbers}
over $\mathbb{K}$.

\begin{definition}
The ring of dual numbers $\hat{\mathbb{K}}$ is the quotient ring defined by
\[
\hat{\mathbb{K}} := \mathbb{K}[\varepsilon]/(\varepsilon^2) = 
\{ p_0 + p_1 \varepsilon \mid p_0,p_1\in \mathbb{K} \}.
\]
We often denote an element in $\hat{\mathbb{K}}$ by 
a symbol with hat such as $\hat{p}$.
\end{definition}
The addition and the multiplication of two dual numbers
are given as
\begin{eqnarray*}
(p_0 + p_1 \varepsilon) + (q_0 + q_1 \varepsilon) &=& (p_0+q_0) + (p_1+q_1) \varepsilon, \\
(p_0 + p_1 \varepsilon) (q_0 + q_1 \varepsilon) &=& (p_0q_0) + (p_1q_0 + p_0q_1) \varepsilon.
\end{eqnarray*}

The following involution is considered to be the dual version of conjugation
\[
 \widetilde{p_0 + p_1 \varepsilon} := p^{*}_0 - p^{*}_1\varepsilon,  
\]
where $p^{*}_i$ is the usual conjugation of $p_{i}$ in $\mathbb{K}$.
(Note that in some literatures $\tilde{\hat{p}}$ is denoted by $\overline{\hat{p}^{*}}$.)

The {\em unit dual numbers} are of special importance.
\begin{definition}
 Let $|\hat{p}|= \sqrt{\hat{p}\tilde{\hat{p}}}$ for $\hat{p}\in \hat{\mathbb{K}}$.
 The unit dual numbers is defined as
 \[
 \hat{\mathbb{K}}_1 := \{ \hat{p} \in \hat{\mathbb{K}} \mid |\hat{p}|=1 \} \subset \hat{\mathbb{K}}.
 \] 
 $\hat{\mathbb{K}}_1$ acts on $\hat{\mathbb{K}}$ by conjugation action
 \[
  \hat{p} \diamond \hat{q} := \hat{p} \hat{q} \tilde{\hat{p}}
 \]
 where $\hat{p} \in \hat{\mathbb{K}}_1, \hat{q}\in \hat{\mathbb{K}}$. 
\end{definition}

The unit dual quaternion $\hat{\H}_1$ is successfully used for skinning in \cite{DLB};
a vector $v=(x,y,z)\in \R^3$ is identified with $1+(xi+yj+zk)\varepsilon \in \hat{\H}_{1}$ and the conjugation action
of $\hat{\H}_1$ preserves the embedded $\R^3$ and its Euclidean metric.
In fact, the conjugation action induces the double cover $\hat{\H}_1 \to \SE(3)$.

On the other hand, when $\mathbb{K}=\R$ or $\C$, the conjugation action is trivial since the 
corresponding dual numbers are commutative.
Therefore, we define the {\em anti-commutative dual complex numbers} (DCN, for short) $\check{\C}$ 
by modifying the multiplication of $\hat{\C}$. That is,
$\check{\C}=\hat{\C}$ as a set, and the algebraic operations are replaced by
\begin{eqnarray*}
(p_0 + p_1 \varepsilon) (q_0 + q_1 \varepsilon) &=& (p_0q_0) + (p_1\tilde{q}_0 + p_0q_1) \varepsilon, \\
\widetilde{p_0 + p_1 \varepsilon} &=& \tilde{p}_0 + p_1\varepsilon, \\
|p_0 + p_1 \varepsilon| &=& |p_0|.
\end{eqnarray*}
The addition and the conjugation action are kept unchanged. Then, 
\begin{theorem}
$\check{\C}$ satisfies distributive and associative laws, and hence, 
has a (non-commutative) ring structure.
\end{theorem}

Similarly to the unit dual quaternion numbers, the unit anti-commutative complex numbers are of particular importance:
\[
\check{\C}_1 := \{\hat{p}\in \check{\C} \mid |\hat{p}|=1\}=\{e^{i\theta}+p_1\varepsilon \in \check{\C} \mid \theta\in \R, p_1\in \C\}.
\]
It forms a group with inverse 
\[
( e^{i\theta}+ p_1\varepsilon)^{-1}=e^{-i\theta}-p_1\varepsilon.
\]
We define an action of $\check{\C}_1$ on $\check{\C}$ by the conjugation.
Now, we regard $\C=\R^2$ as usual.
Identifying $v \in \C$ with $1+v\varepsilon \in \check{\C}$, 
we see that $\check{\C}_1$ acts on $\C$ as rigid transformation.

\section{Relation to $\SE(2)$}
In the previous section, we constructed the unit anti-commutative dual complex numbers
$\check{\C}_1$ and its action on $\C=\R^2$ as rigid transformation.
Recall that $\hat{p}=p_0 + p_1\varepsilon \in \check{\C}_1$ acts on $v\in \C$ by
\begin{equation}\label{action}
 \hat{p} \diamond (1+v\varepsilon) = 
 (p_0+p_1\varepsilon) (1+v\varepsilon) (\tilde{p}_0+p_1\varepsilon)
 = 1 + (p_0^2v + 2p_0p_1)\varepsilon,
\end{equation}
that is, $v$ maps to $p_0^2v + 2p_0p_1$.
For example, when $p_1=0$, $v\in \C$ is mapped to $p_0^2 v$, which 
is the rotation around the origin of degree $2\arg(p_0)$ since $|p_0|=1$.
On the other hand, when $p_0=1$, the action corresponds to the translation by $2p_1$.
In general, we have 
\begin{align*}
 \varphi: \check{\C}_1 & \to \SE(2) \\
 p_0 + p_1\varepsilon & \mapsto 
 \begin{pmatrix} \Re(p_0^{2}) & -\Im(p_0^{2}) & \Re(2p_0p_1) \\
 \Im(p_0^{2}) & \Re(p_0^{2}) & \Im(2p_0p_1) \\
  0 & 0 & 1
 \end{pmatrix},
\end{align*}
where $\Re(2p_0p_1)$ (respectively, $\Im(2p_0p_1)$) is the real (respectively, imaginary)
 part of $2p_0p_1\in \C$.
 Note that this gives a surjective group homomorphism
 $\varphi: \check{\C}_1 \to \SE(2)$ whose kernel is $\{\pm 1\}$.
 That is, 
 the preimage of any 2D rigid transformation consists of exactly two unit DCN's with opposite signs.
\begin{example}
We compute the DCN's $\pm \hat{p}\in \check{\C}_1$ which represent
$\theta$-rotation around  $v\in \C$.
It is the composition of the following three rigid transformations:
$(-v)$-translation, $\theta$-rotation around the origin, and $v$-translation.
Thus, we have
\[\hat{p}= \left( 1+\dfrac{v}{2}\varepsilon \right) \cdot 
\left(\pm e^{\frac{\theta}{2}i}\right) \cdot \left(1-\dfrac{v}{2}\varepsilon \right)
=\pm \left(e^{\frac{\theta}{2}i}+ \left(e^{-\frac{\theta}{2}i}-e^{\frac{\theta}{2}i} \right)\dfrac{v}{2}\varepsilon \right).
 \]
\end{example}

\section{Relation to the dual quaternion numbers}\label{dan}
The following ring homomorphism
\[
 \check{\C} \ni p_0 + p_1\varepsilon \mapsto p_0 + p_1 j\varepsilon \in \hat{\H}
\]
is compatible with the involution and the conjugation, and preserves the norm.
Furthermore, if we identify $v=(x,y)\in \R^2$ with $1+(xj+yk)\varepsilon \ (=1+(x+yi)j\varepsilon)$, the above map commutes with the action.
From this point of view, DCN is 
nothing but a sub-ring of DQN.

Note also that $\check{\C}$ can be embedded in the ring
of the $2\times 2$-complex matrices by
\[
p_0 + p_1 \varepsilon \mapsto \begin{pmatrix} p_0 & p_1 \\ 0 & \tilde{p}_0 \end{pmatrix}.
\]
Then 
\[
 \widetilde{ \begin{pmatrix} p_0 & p_1 \\ 0 & \tilde{p}_0 \end{pmatrix}}=
  \begin{pmatrix} \tilde{p}_0 & p_1 \\ 0 & p_0 \end{pmatrix},  \qquad
  \left|  \begin{pmatrix} p_0 & p_1 \\ 0 & \tilde{p}_0 \end{pmatrix} \right|^2 = 
  \det\begin{pmatrix} p_0 & p_1 \\ 0 & \tilde{p}_0 \end{pmatrix}.
\]

We thus have various equivalent presentations of DCN.
However, our presentation of $\check{\C}$ as the anti-commutative dual numbers
 is easier to implement and more efficient.


\section{Interpolation of 2D rigid transformations}\label{blend}
First, recall that for the positive real numbers $x,y\in \R_{> 0}$, 
there are two typical interpolation methods:
\[
 (1-t)x + t y, \quad t \in \R
\]
and
\[
 (yx^{-1})^t x, \quad t\in \R.
\]

The first method can be generalized to DQN as 
the {\em Dual quaternion Linear Blending} in \cite{DLB}.
Similarly, a DCN version of {\em Dual number Linear Blending (DLB, for short)} is given as follows:
\begin{definition}
For $\hat{p}_1, \hat{p}_2, \ldots, \hat{p}_n\in \check{\C}_1$, we define
\begin{equation}\label{eq:DLB}
 DLB(\hat{p}_1, \hat{p}_2, \ldots, \hat{p}_n; w_1, w_2, \ldots, w_n)=
 \dfrac{w_1\hat{p}_1+w_1\hat{p}_2+\cdots+w_n\hat{p}_n}{|w_1\hat{p}_1+w_1\hat{p}_2+\cdots+w_n\hat{p}_n|},
\end{equation}
where $w_1, w_2, \ldots, w_n\in \R$.
Note that the denominator can become $0$ and
for those set of $w_{i}$'s and $\hat{p}_{i}$'s DLB cannot be defined.
\end{definition}
A significant feature of DLB is that it is distributive
(it is called {\em bi-invariant} in some literatures). That is,
the following holds:
\[
 \hat{p}_0 DLB(\hat{p}_1, \hat{p}_2, \ldots, \hat{p}_n; w_1, w_2, \ldots, w_n)=
  DLB(\hat{p}_0\hat{p}_1, \hat{p}_0\hat{p}_2, \ldots, \hat{p}_0\hat{p}_n; w_1, w_2, \ldots, w_n).
\]
This property is particularly important when transformations are given in a certain hierarchy such as
in the case of skinning;
if the transformation assigned to the root joint is modified, 
the skin associated to lower nodes is deformed consistently.

Next, we consider the interpolation of the second type.
For this, we need the exponential and the logarithm maps for DCN.
\begin{definition}
For $\hat{p}=p_0+p_1\varepsilon\in \check{\C}$, we define
\[
\exp{\hat{p}}=\sum_{n=0}^\infty \dfrac{(p_0+p_1\varepsilon)^n}{n!}=e^{p_0}+\dfrac{(e^{p_0}-e^{\tilde{p}_0})}{p_0-\tilde{p}_0}p_1\varepsilon.
\]
\end{definition}
When $\exp{\hat{p}}\in \check{\C}_1$,
we can write $p_0 = \theta i$ for some $ -\pi \le \theta < \pi$, and 
\[
\exp{\hat{p}}= e^{\theta i}+ \dfrac{\sin\theta}{\theta}p_1\varepsilon.
\]
For $\hat{q}=e^{\theta i}+q_1\varepsilon\in \check{\C}_1$, we define
\[
 \log(\hat{q})=\theta i +\dfrac{\theta}{\sin\theta}q_1 \varepsilon.
\]
As usual, we have $\exp(\log(\hat{q}))=\hat{q}$ and $\log(\exp(\hat{p}))=\hat{p}$.
Note that this gives the following {\em Lie correspondence} (see \cite{lie,AO1,AO2})
\begin{align*}
 \exp: \dcn & \to \check{\C}_1, \\
 \log: \check{\C}_1 & \to \dcn,
\end{align*}
where $\dcn=\{ \theta i + p_1 \varepsilon \in \check{\C} \mid \theta \in \R, p_1\in \C \}\simeq \R^3$. 
We have the following commutative diagram:
\[
\xymatrix{
\dcn \ar[r]^{d\varphi} \ar[d]^{\exp} & \se(2) \ar[d]^\exp \\
\check{\C}_1 \ar[r]^\varphi & \SE(2),
}
\]
where 
\begin{align*}
 d\varphi: \dcn & \to \se(2) \\
 \theta i + (x+yi)\varepsilon &\mapsto \begin{pmatrix} 0 & -2\theta & 2x \\ 2\theta & 0 & 2y \\ 0 & 0 & 0 \end{pmatrix},
\end{align*}
is an isomorphism of $\R$-vector spaces.

The following is a DCN version of SLERP \cite{SLERP}.
\begin{definition}
 SLERP interpolation from $\hat{p}\in \check{\C}_1$ to $\hat{q}\in \check{\C}_1$ is given by
 \[
  \mathrm{SLERP}(\hat{p},\hat{q};t) = (\hat{q}\hat{p}^{-1})^t\hat{p}=
  \exp(t\log(\hat{q}\hat{p}^{-1}))\hat{p},
 \]
 where $t\in \R$.
\end{definition}
This gives a uniform angular velocity interpolation of two DCN's,
while DLB can blend three or more DCN's without the uniform angular velocity property.

\section{Computational cost}\label{implementation}
We compare the following four methods for 2D rigid transformation:
our DCN, the unit dual quaternion numbers (see \S \ref{dan}), 
the $3\times 3$-real (homogeneous) matrix representation of $\SE(2)$,
and the $2\times 2$-complex matrices described below.

We list the computational cost in terms of the number of floating point operations for 
\begin{itemize}
\item transforming a point
\item composing two transformations
\item converting a transformation to the standard $3\times 3$-real matrix representation
(except for the $3\times 3$-real matrix case where it shows the computational cost to convert to DCN representation).
\end{itemize}
We also give the memory usage for each presentation in terms of the number of floating point units necessary to store a transformation.

\begin{table}[ht]
\begin{tabular}{c|c|c|c|c}
 & transformation & composition & conversion & memory usage\\
 \hline
 DCN 					& 22  FLOPs 	& 20 FLOPs	& 	15 FLOPs	& 4 scalars \\
 DQN 					& 92 FLOPs	& 88 FLOPs	& NA		& 8 scalars \\
 $2\times 2$-complex matrix 	& 112 FLOPs		& 56 FLOPs	& 15  FLOPs	& 8 scalars  \\
 $3\times 3$-real matrix 		& 15 FLOPs 		& 45 FLOPs 	& 18	FLOPs(to DCN) 	& 9 scalars
\end{tabular}
\caption{Comparison of computational cost}
\end{table}

Note that when a particular application requires to apply a single transformation to a lot of points,
it is faster to first convert the DCN to a $3\times 3$-real matrix.

\section{A C++ Library}
We implemented our DCN in a form of C++ library. 
Though it is written in C++, it should be easy to translate to any language.
You can download the MIT-licensed source code at \cite{source}.
We also developed a small demo application called 
the 2D probe-based deformer (\cite{deformer}) for tablet devices running OpenGL ES
(OpenGL for Embedded Systems).
Thanks to the efficiency of DCN and 
touch interface, 
it offers interactive  and intuitive operation.
Although we did not try, we believe DCN works well with 2D skinning just as DQN does with 3D skinning.

\begin{figure}[htbp]
 \begin{center}
  \includegraphics[height=5cm]{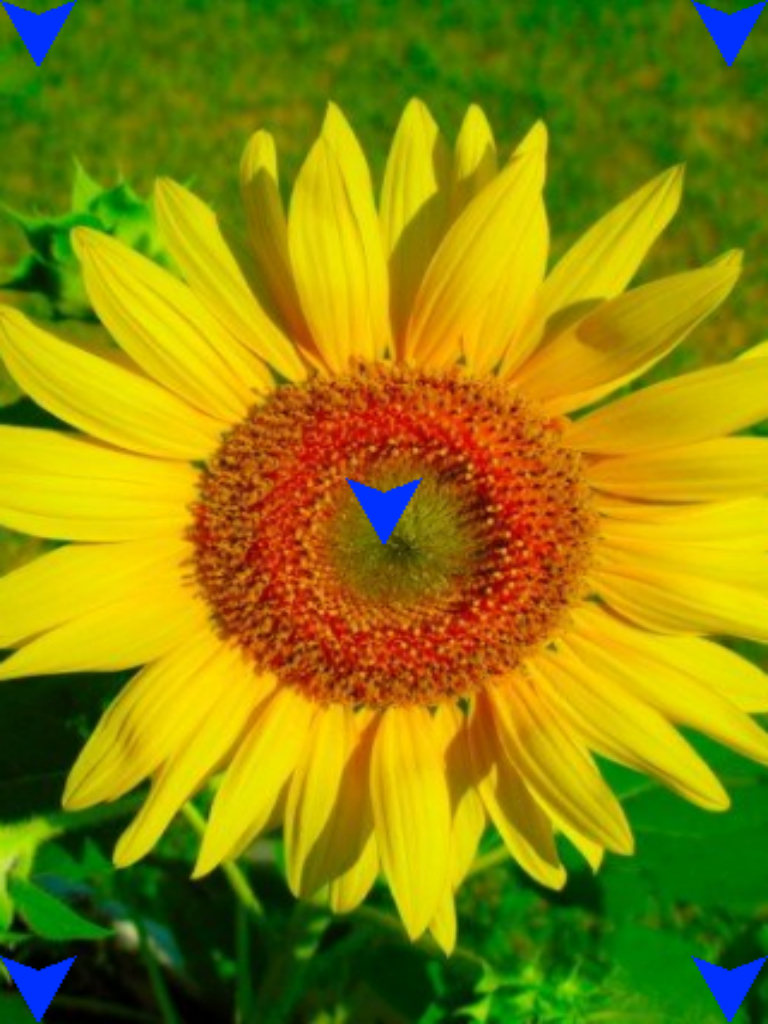} \hspace{5mm}
  \includegraphics[height=5cm]{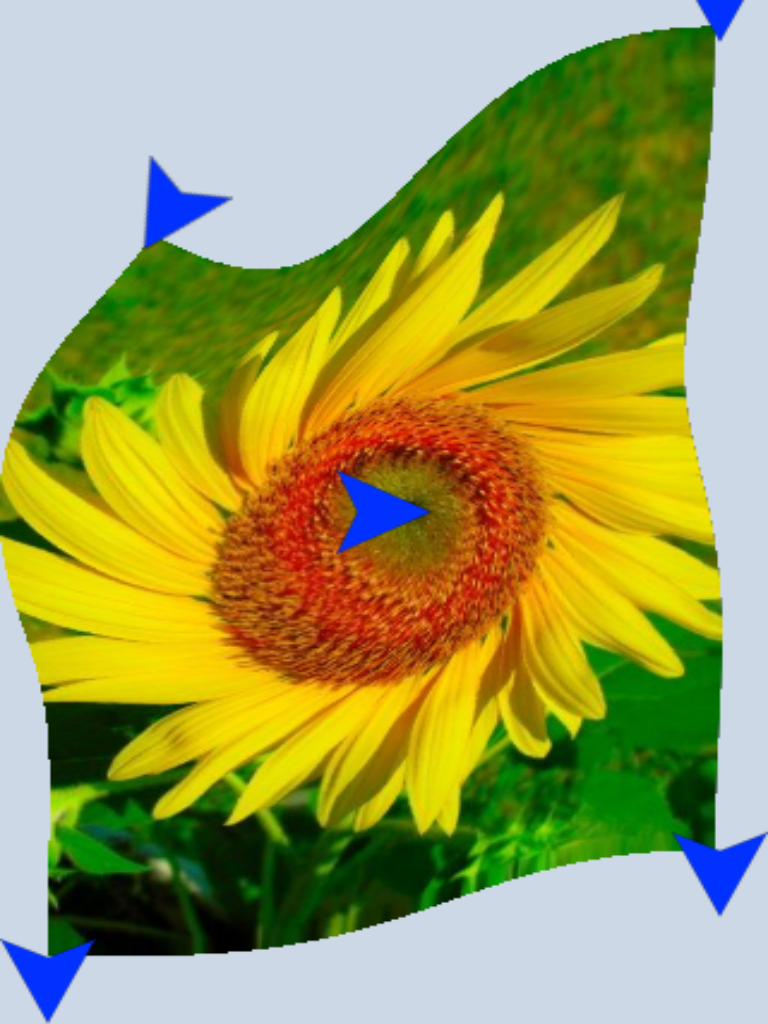} \\ \vspace{5mm}
  \includegraphics[height=5cm]{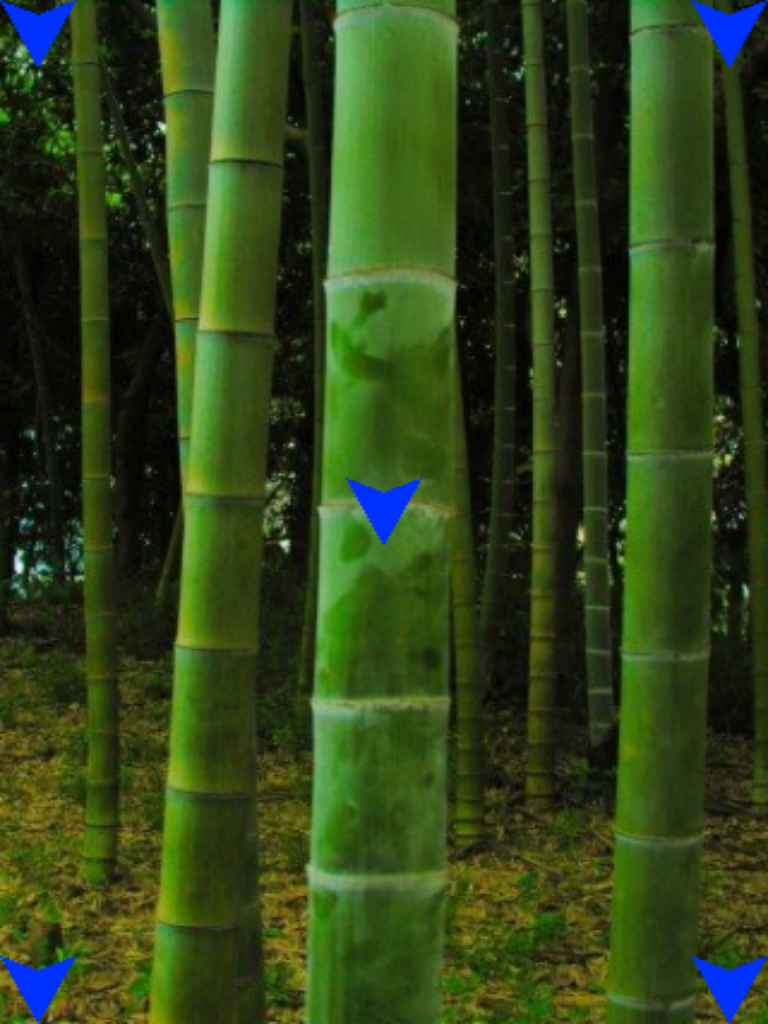}\hspace{5mm}
  \includegraphics[height=5cm]{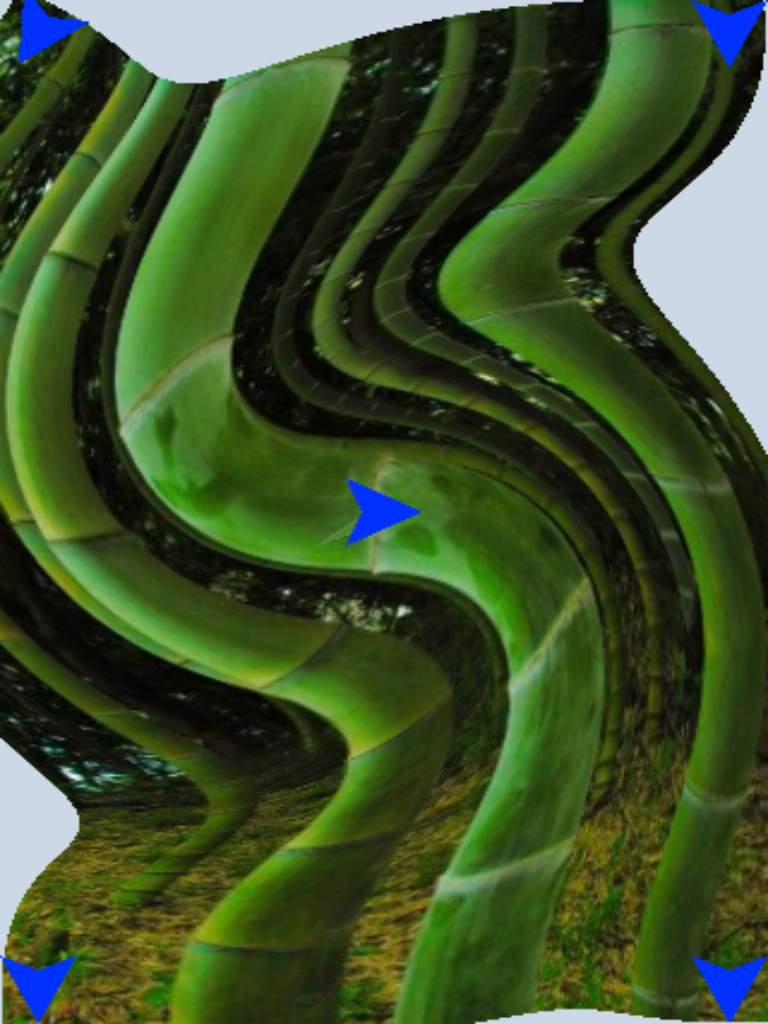}
 \end{center}
 \caption{Left: set up initial positions of (blue) probes.
 Right: the target picture is deformed according to user's action on the probes.}
\end{figure}

Here we briefly discuss the algorithm of the application.
One can place an arbitrary number of {\em probes} $P_{1},\ldots,P_{n}$ on a target image.
The target image is represented by a textured square mesh with vertices $v_{1},\ldots,v_{m}$.
The weight $w_{ij}$ of $P_{i}$'s effect on $v_{j}$ is painted by user or automatically calculated 
from the distance between $v_{j}$ and $P_{i}$. 
When the probes are rotated or translated by the user's touch gesture,
the DCN $\hat{p}_{i}$ is computed which maps the initial state of $P_{i}$ to its current state.
The vertex $v_{j}$ is transformed by
\[
 v_{j} \mapsto DLB(\hat{p}_1, \hat{p}_2, \ldots, \hat{p}_n; w_{1j}, w_{2j}, \ldots, w_{nj}) \diamond v_{j}.
\]
(See Equations (\ref{action}) and (\ref{eq:DLB}) respectively for the definition of the action $\diamond$ and DLB.)

\section*{Acknowledgements}
This work was supported by Core Research for Evolutional Science and Technology (CREST) Program 
``Mathematics for Computer Graphics'' of Japan Science and Technology Agency (JST).
The authors are grateful for S. Hirose at OLM Digital Inc.,
and Y. Mizoguchi, S. Yokoyama, H. Hamada, and K. Matsushita at Kyushu University
for their valuable comments.

\end{document}